

\font\titlefont = cmr10 scaled\magstep 4
\font\sectionfont = cmr10
\font\littlefont = cmr5
\font\eightrm = cmr8

\def\ss{\scriptstyle}
\def\sss{\scriptscriptstyle}

\newcount\tcflag
\tcflag = 0  

\ifnum\tcflag = 0 \magnification = 1200 \fi  

\global\baselineskip = 1.2\baselineskip
\global\parskip = 4pt plus 0.3pt
\global\abovedisplayskip = 18pt plus3pt minus9pt
\global\belowdisplayskip = 18pt plus3pt minus9pt
\global\abovedisplayshortskip = 6pt plus3pt
\global\belowdisplayshortskip = 6pt plus3pt


\def\endignore{}
\def\ignore #1\endignore{}

\newcount\dflag
\dflag = 0


\def\monthname{\ifcase\month
\or January \or February \or March \or April \or May \or June%
\or July \or August \or September \or October \or November %
\or December
\fi}

\newcount\dummy
\newcount\minute  
\newcount\hour
\newcount\localtime
\newcount\localday
\localtime = \time
\localday = \day

\def\advanceclock#1#2{ 
\dummy = #1
\multiply\dummy by 60
\advance\dummy by #2
\advance\localtime by \dummy
\ifnum\localtime > 1440 
\advance\localtime by -1440
\advance\localday by 1
\fi}

\def\settime{{\dummy = \localtime%
\divide\dummy by 60%
\hour = \dummy
\minute = \localtime%
\multiply\dummy by 60%
\advance\minute by -\dummy
\ifnum\minute < 10
\xdef\spacer{0} 
\else \xdef\spacer{}
\fi %
\ifnum\hour < 12
\xdef\ampm{a.m.} 
\else
\xdef\ampm{p.m.} 
\advance\hour by -12 %
\fi %
\ifnum\hour = 0 \hour = 12 \fi
\xdef\timestring{\number\hour : \spacer \number\minute%
\thinspace \ampm}}}



\def\endtitle{}
\def\title#1\endtitle{\vskip.5in\titlefont
\global\baselineskip = 2\baselineskip
#1\vskip.4in
\baselineskip = 0.5\baselineskip\rm}

\def\endauthors{}
\def\authors#1\endauthors{#1}

\def\endabstract{}
\def\abstract#1\endabstract{\vskip .3in%
\centerline{\sectionfont\bf Abstract}%
\vskip .1in
\noindent#1}

\def\nopageonenumber{\footline={\ifnum\pageno<2\hfil\else
\hss\tenrm\folio\hss\fi}}  

\newcount\nsection
\newcount\nsubsection

\def\section#1{\global\advance\nsection by 1
\nsubsection=0
\bigskip\noindent\centerline{\sectionfont \bf \number\nsection.\ #1}
\bigskip\rm\nobreak}

\def\subsection#1{\global\advance\nsubsection by 1
\bigskip\noindent\sectionfont \sl \number\nsection.\number\nsubsection)\
#1\bigskip\rm\nobreak}


\def\appendix#1#2{\bigskip\noindent%
\centerline{\sectionfont \bf Appendix #1.\ #2}
\bigskip\rm\nobreak}


\newcount\nref
\global\nref = 1

\def\therefs{}


\def\ref#1#2{\xdef #1{[\number\nref]}
\ifnum\nref = 1\global\xdef\therefs{\item{[\number\nref]} #2\ }
\else
\global\xdef\oldrefs{\therefs}
\global\xdef\therefs{\oldrefs\vskip.1in\item{[\number\nref]} #2\ }%
\fi%
\global\advance\nref by 1
}

\def\listrefs{\vfill\eject\section{References}\therefs}


\newcount\nfoot
\global\nfoot = 1

\def\foot#1#2{\xdef #1{(\number\nfoot)}
\footnote{${}^{\number\nfoot}$}{\eightrm #2}
\global\advance\nfoot by 1
}


\newcount\nfig
\global\nfig = 1
\def\thefigs{} 

\def\figure#1#2{\xdef #1{(\number\nfig)}
\ifnum\nfig = 1\global\xdef\thefigs{\item{(\number\nfig)} #2\ }
\else
\global\xdef\oldfigs{\thefigs}
\global\xdef\thefigs{\oldfigs\vskip.1in\item{(\number\nfig)} #2\ }%
\fi%
\global\advance\nfig by 1 } 

\def\fig#1{\xdef #1{(\number\nfig)}
\global\advance\nfig by 1 } 


\newcount\cflag
\newcount\nequation
\global\nequation = 1
\def\eqlabel{(1)}

\def\nexteqno{\ifnum\cflag = 0
\global\advance\nequation by 1
\fi
\global\cflag = 0
\xdef\eqlabel{(\number\nequation)}}

\def\lasteqno{\global\advance\nequation by -1
\xdef\eqlabel{(\number\nequation)}}

\def\label#1{\xdef #1{(\number\nequation)}
\ifnum\dflag = 1
{\escapechar = -1
\xdef\draftname{\littlefont\string#1}}
\fi}

\def\clabel#1#2{\xdef\eqlabel{(\number\nequation #2)}
\global\cflag = 1
\xdef #1{\eqlabel}
\ifnum\dflag = 1
{\escapechar = -1
\xdef\draftname{\string#1}}
\fi}

\def\cclabel#1#2{\xdef\eqlabel{#2)}
\global\cflag = 1
\xdef #1{\eqlabel}
\ifnum\dflag = 1
{\escapechar = -1
\xdef\draftname{\string#1}}
\fi}


\def\eeq{}

\def\eqnn #1\eeq{$$ #1 $$}

\def\eq #1\eeq{
\ifnum\dflag = 0
{\xdef\draftname{\ }}
\fi 
$$ #1
\eqno{\eqlabel \rlap{\ \draftname}} $$
\nexteqno}



\def\eol{& \eqlabel \rlap{\ \draftname} \crcr
\nexteqno
\xdef\draftname{\ }}

\def\eeol{& \eqlabel \rlap{\ \draftname}
\nexteqno
\xdef\draftname{\ }}

\def\eolnn{\cr
\global\cflag = 0
\xdef\draftname{\ }}

\def\eeolnn{\xdef\draftname{\ }}

\def\eqa #1\eeq{
\ifnum\dflag = 0
{\xdef\draftname{\ }}
\fi 
$$ \eqalignno{ #1 } $$
\global\cflag = 0}


\def\ie{{\it i.e.\/}}


\def\cmp#1#2#3{{\it Comm.\ Math.\ Phys.} {\bf #1} (19#2) #3}

\def\npb#1#2#3{{\it Nucl.\ Phys.} {\bf B#1} (19#2) #3}

\def\prd#1#2#3{{\it Phys.\ Rev.} {\bf D#1} (19#2) #3}

\def\prep#1#2#3{{\it Phys.\ Rep.} {\bf C#1} (19#2) #3}


\global\nulldelimiterspace = 0pt



\def\frac#1#2{{{#1} \over {#2}}\,}  
\def\hf{{1\over 2}}
\def\nth#1{{1\over #1}}

\def\ppartial#1#2{{{\partial #1} \over {\partial #2}}}  

\def\Square{{\vbox {\hrule height 0.6pt\hbox{\vrule width 0.6pt\hskip 3pt
        \vbox{\vskip 6pt}\hskip 3pt \vrule width 0.6pt}\hrule height 0.6pt}}}
\def\Dsl{\hbox{/\kern-.6700em\it D}} 
\def\dsl{\hbox{/\kern-.5300em$\partial$}}
\def\pxpsl{\hbox{/\kern-.5600em$p$}}
\def\ssl{\hbox{/\kern-.5300em$s$}}
\def\epssl{\hbox{/\kern-.5100em$\epsilon$}}
\def\delsl{\hbox{/\kern-.6300em$\nabla$}}
\def\lxpsl{\hbox{/\kern-.4300em$l$}}
\def\elxpsl{\hbox{/\kern-.4500em$\ell$}}
\def\kxpsl{\hbox{/\kern-.5100em$k$}}
\def\qxpsl{\hbox{/\kern-.5000em$q$}}
\def\sla#1{\raise.15ex\hbox{$/$}\kern-.57em #1}



\def\roughly#1{\mathrel{\raise.3ex\hbox{$#1$\kern-.75em\lower1ex\hbox{$\sim$}}}}

\def\gsim{\roughly>}





\def\Scf{{\cal F}}

\def\Scr{{\cal R}}

\def\Sct{{\cal T}}


\def\ssh{{\sss H}}

\def\ssl{{\sss L}}




\def\avg#1{\langle #1 \rangle}






\def\th{T_{\sss H}}

\def\lh{L_{\sss H}}
\def\rs{r_s}

\def\lrderiv#1{{{\buildrel\leftrightarrow\over{\partial}}_{#1}}}


\line{hep-th/9510159 \hfill McGill-95/50}
\rightline{October, 1995.}
\vskip .2in

\title
\centerline{Hawking Radiation}
\centerline{and Ultraviolet Regulators}
\endtitle

\authors
\centerline{N.~Hambli and C.P.~Burgess}
\vskip .15in
\centerline{\it Physics Department, McGill University}
\centerline{\it 3600 University St., Montr\'eal, Qu\'ebec, CANADA,
H3A 2T8.}
\endauthors

\abstract
Polchinski has argued that the prediction of Hawking radiation must be
independent of the details of unknown high-energy physics because the
calculation may be performed using `nice slices', for which the adiabatic
theorem may be used. If this is so, then any calculation using a manifestly
covariant --- and so slice-independent --- ultraviolet regularization must
reproduce the standard Hawking result. We investigate the dependence of the
Hawking radiation on such a short-distance regulator by calculating it using a
Pauli--Villars regularization scheme. We find that the regulator scale,
$\Lambda$,  only contributes to the Hawking flux by an amount that is
exponentially small in the large variable ${\Lambda}/{T_\ssh} \gg 1$,
where $T_\ssh$ is the Hawking temperature; in agreement with Polchinski's
arguments. We also solve a technical puzzle concerning the relation between the
short-distance singularities of the propagator and the Hawking effect.
\endabstract


\section{Introduction}

The prediction that very massive stars must end their days as black holes has
by now become deeply ingrained into common astrophysical lore. Our belief in
this result rests in no small part on the continued success with
which General Relativity accounts for observations, both within the solar
system and beyond.

Part of the progress of the last twenty years has been the
integration of this success into the broader body of laws which describe the
other known, nongravitational, interactions. It is now understood that, in
spite of the notorious obstacles to constructing a full quantum theory of
gravity, semiclassical General Relativity can be interpreted as a controllable
low-energy approximation to whatever unknown physics might
ultimately describe nature on the very shortest of length scales. In this
sense, General Relativity joins other venerable, but nonrenormalizable,
low-energy effective   theories
\ref\effective{See, for example, S. Weinberg, Physica {\bf 96A}
(1979) 327;  J.
Polchinski, \npb{231}{84}{269};  J.~Polchinski, in the proceedings
of TASI 1992
(Boulder Colorado);
J. F. Donoghue, \prd{50}{94}{3874}.}
\effective, and semiclassical calculations are justified for observables
that vary on distance scales that are long compared to the Planck length.

Perhaps the biggest surprise to emerge from the study of semiclassical quantum
physics in the presence of macroscopic gravitational fields is Hawking's
discovery
\ref\hawking{S.W.~Hawking, {\it Nature} {\bf 248} (1974) 30;
\cmp{43}{75}{199}.}
\hawking, that black holes constantly radiate subatomic particles. These
particles dominantly emerge far from the hole with energies that
are of order the Hawking temperature: $E \simeq \th \equiv (4 \pi \rs)^{-1}$,
where\foot\units{We use fundamental units, $\ss \hbar = c = k_{\sss
B} = 1$, throughout.} $\rs = 2 G M$ is the Schwarzschild radius for a black
hole of mass $M$. Provided that the hole is sufficiently massive --- $G M^2 \gg
1$ (or, in cgs units, $M \gg 22 \; \mu g$) --- the radiation is a
long-wavelength effect, and so one expects its semiclassical description to be
justified.

It therefore comes as something of a surprise, as was originally
emphasized in
\ref\unruh{W.G.~Unruh, \prd{14}{76}{870}, \prd{15}{77}{365}.}
\unruh, and more recently in
\ref\tedone{T.~Jacobson, \prd{44}{91}{1731}.}
\ref\tedtwo{T.~Jacobson, \prd{48}{93}{728 }.}
\tedone\ and \tedtwo, to find that the standard derivations of the
Hawking effect (in four dimensions) make reference in one way or another to
physics at extremely short distances. This is true both of Hawking's original
derivation, as well as of more modern alternatives
\ref\fh{K.~Fredenhagen and R.~Haag, \cmp{127}{90}{273}.}
\fh.

The short distances arise because the Hawking radiation is defined
to be the flux which emerges at very late times, well after all of the
transients associated  with the stellar collapse itself have passed. However,
in the usual derivations the radiation which emerges at the Hawking temperature
at such late times is strongly redshifted as it climbs out of the
black hole's gravitational well. Alternatively, in the formalism set up in
Ref.~\fh, the outgoing flux is derived from the short-distance form for the
radiated particle's two-point (Hadamard) function (see below for details) as
its position arguments, $x$ and $x'$, approach one another and the event
horizon.

Polchinski
\ref\niceslice{
D.~A.~Lowe, J.~Polchinski, L.~Susskind,
L.~Thorlacius and J.~Uglum, {\it Black Hole
Complementarity {\it vs.} Locality},
{\bf hep-th/9506138}; J.~Polchinski, {\it ``String Theory And Black Hole
Complementarity''}, {\bf hep-th/9507094}.}
\niceslice, on the other hand, has argued persuasively that, in
spite of these appearances, Hawking radiation is nevertheless a
robust feature of the long-distance theory. His arguments use the ability,
in principle,  to perform one's calculations using only `nice slices' for
which curvatures are everywhere small, and for which the adiabatic
theorem ensures all high-frequency modes must be in their ground
state.

Our purpose here is to present evidence supporting Polchinski's arguments using
an explicit calculation of the Hawking radiation in a simple model. We perform
the following consistency check on these arguments: if the existence of nice
slices guarantees that the Hawking flux is independent of the details of
short-distance physics, then any reasonable manifestly-covariant --- and so
slice-independent --- ultraviolet regularization must also not affect this
flux. We test this by computing the Hawking radiation using a minimally-coupled
massive scalar field in the presence of a Schwarzschild black hole,
using a Pauli--Villars ultraviolet regularization. We are able to implement
this
regularization by suitably adapting the methods of Fredenhagen and Haag
Ref.~\fh. We find that all of the cutoff dependence vanishes exponentially in
the limit $\Lambda \gg \th$, in agreement with Polchinski's arguments.

The details of this calculation are described in Section (2). In Section
(3), we resolve an apparent paradox concerning the relation between the Hawking
radiation and the absence of short-distance singularities of the two-point
function in the regulated theory. Our conclusions are  summarized
in Section (4).

\section{A Regulated Example}

In this section we compute the dependence of the Hawking flux on the
short-distance regulator.

We take as our observable the outgoing energy flux per unit time,
$\Scf \equiv -\avg{{T_t}^r}$, as seen at very late times and at  a
very large distance from the black hole, with the
average taken in the state which corresponds to the vacuum at very early
times before the black hole has formed. $t$ and $r$ here represent
the usual
Schwarzschild coordinates, in terms of which $ds^2 = - \left(1 -
\rs/r \right)
\, dt^2 + \left(1 - \rs/r \right)^{-1} \, dr^2 + r^2 \, (d\theta^2
+ \sin^2
\theta \, d \phi^2)$. $\Scf$ is related to the total black hole
luminosity by
$\lh = \int \Scf \, r^2 \sin\theta \, d\theta d\phi$.

For a minimally-coupled scalar field the stress tensor is quadratic
in the
field operator, and so its expectation may be expressed in terms of the
coincidence limit of the Hadamard two-point function: $G(x,x') \equiv \hf
\avg{ \varphi(x) \varphi(x') + \varphi(x') \varphi(x) }$. In our case:
\label\coincidence
\eqa  \Scf  &\equiv  - \avg{ {T_t}^r} = - \avg{ T_{tr^\star}} \eolnn
&= - \hf  \lim_{x' \to x} \left( \ppartial{}{t'} \,\ppartial{}{r^\star} +
\ppartial{}{{r^\star}'}\, \ppartial{}{t} \right) \; G(x,x'), \eeol \eeq
where $r^\star$ is the `tortoise' coordinate: $r^\star \equiv r  +
\rs \, \ln \left[
(r/\rs)  -1 \right]$. The problem reduces to the calculation of
$G(x,x')$.

\subsection{The Regulator}

Since $G(x,x')$ is singular as $x' \to x$, the components of the
stress tensor
usually diverge, and so must be regularized and renormalized.
(Off-diagonal
components are typically finite in Schwarzschild, however.) We choose to
perform this regularization {\`a} la Pauli-Villars --- \ie\ by
introducing
additional  fields, $\varphi_i(x)$, some with the `wrong'-sign
kinetic energies,
in such a  way as to ensure the finiteness of the coincidence limit of
\label\regg
\eq  G_{\rm reg}(x,x')  = \sum_i  \epsilon_i G_i(x,x'), \eeq
where $\epsilon_i = \pm$ keeps track of the sign of the
corresponding field's
kinetic energy. We should also point out that the sum
in eq.~\regg\ includes the contribution from the
physical field of mass $m$ whose $\epsilon = +$.
Our purpose  is ultimately to determine how  $\Scf$
depends on
the masses of the regularization fields, $M_i$, in the limit that
$M_i \sim
\Lambda \gg \th \gsim m$, where $\Lambda$ is the inverse of a
covariantly-defined cutoff length (see below), and $m$ is the mass of the
original scalar field.

The properties of the regulator fields that are required may be directly
calculated from the known divergence structure for minimally-coupled
free scalar fields  propagating through macroscopic background
fields. For a
scalar field of mass, $m$, there are three independent types of
divergences,
which are known to be proportional to the following three coefficients
\ref\gilkey{P.B.~Gilkey, {\it J.~Diff.~Geom.} {\bf 10} (1975) 601;
S.M.~Christensen, \prd{14}{76}{2490}, B.S.~DeWitt,
\prep{19}{75}{296}.}\gilkey:
\ref\weinberg{S.~Weinberg, {\it Gravitation and Cosmology:
Principles and Applications of the General Theory of Relativity} (1982),
(New York: Wiley).}\foot\conventions{These expressions use the
conventions of
Ref.~\weinberg.}
\label\divergences
\eqa  C_0 &\equiv m^4 \; [a_0] =  m^4 ,\eolnn
C_1 &\equiv m^2 \; [a_1] = - {m^2 \over 6} \, R , \qquad \hbox{and}\eol
C_2 &\equiv [a_2] = \nth{180} \, R_{\mu\nu\lambda\rho} \,
R^{\mu\nu\lambda\rho}
- \nth{180} \, R_{\mu\nu} \, R^{\mu\nu} - \nth{30} \, \Square \, R
+ \nth{72} \,
R^2. \eeolnn \eeq %
Cancellation of all of these short-distance singularities amongst the
Pauli-Villars fields is therefore equivalent to the following conditions:
\label\conditions
\eqa  1 + \sum_i  \epsilon_i \,  &= 0,  \eolnn
m^2 + \sum_i \epsilon_i \, M^2_i  &= 0, \eol
m^4 + \sum_i \epsilon_i \, M^4_i  &= 0. \eeolnn \eeq
As is easily verified, a solution to these equations
is given by: $\epsilon_1 = \epsilon_2 = +$, $M^2_1 = M^2_2 = 3
\Lambda^2 + m^2$; $\epsilon_3 = \epsilon_4 = -$, $M^2_3 = M^2_4 =
\Lambda^2 + m^2$; and $\epsilon_5 = -$, $M^2_5 = 4 \Lambda^2 +
m^2$.

\subsection{Computing the Hawking Flux}

We may now use $G_{\rm reg}(x,x')$ to compute the
$\Lambda$-dependence of the Hawking flux. We do so
by adapting the arguments of Ref.~\fh\ to our example,
since  this formulation of the calculation is easily
applied to massive fields.

Starting from the definition of $G_{\rm reg}(x,x')$, and
eq.~\coincidence, we see that the Hawking flux may be simply
written as $\displaystyle{\Scf = \sum_i \epsilon_i \Scf_i}$,
where $\Scf_i$ is the Hawking flux due to a minimally-coupled
scalar field of mass $M_i$. A straightforward application of
the techniques of Ref.~\fh\ gives the usual result:
\label\fluxresult
\eq \Scf_i (M_i) = { 1 \over 4 \pi^2 r^2} \sum_{\ell m} \left|
Y_{\ell m}(\theta,\phi) \right|^2 \; \int_{M_i}^\infty d\omega \;
\left| \Sct_\ell(\omega,M_i) \right|^2 { \omega \over
e^{\omega/\th} -1 }.
\eeq
In this expression, $(r,\theta,\phi)$ are the Schwarzschild coordinates
for the point at which $\Scf_i$ is computed, and $Y_{\ell m}$ are the
usual spherical harmonics. $|\Sct_\ell(\omega,M_i)|^2 \le 1$ is the
probability that an outgoing particle of mass $M_i$ and energy
$\omega$ (as seen by the stationary observers at infinity)
is transmitted from the event horizon ($r^\star \to - \infty$) out to
infinity, rather than being scattered back to the horizon by the black hole's
gravitational field.

Since the regulator fields all satisfy $M_i \gg \th$, it suffices
to use the
asymptotic form for the flux in this limit. In this limit the
frequency integral
may be bounded from above:
\label\upperbounds
\eqa
\int_{M_i}^\infty d\omega \;  \left| \Sct_\ell(\omega,M_i)
\right|^2 { \omega
\over e^{\omega/\th} -1 } & \le 2 \int_{M_i}^\infty d\omega \;  \omega \,
e^{-\omega/\th} \eolnn
& \le 2 \, \th^2 \left(1 + {M_i\over \th} \right) \; e^{-
M_i/\th}. \eeol \eeq
We see that, for $\Scf_i$, every term in the sum over $\ell$ is
exponentially
small in $\Lambda/\th$. Of course, this is just what would be
expected for a
thermal radiation spectrum.

One might worry that, although each term in the sum over $\ell$ is
exponentially small, it may be that the series sums to a result
which is {\it
not} exponentially suppressed. This does not happen, however,
because a much
stronger bound is possible for $|\Sct_\ell|^2$ when  $\ell$ becomes
sufficiently
large. The better bound arises because for large angular momenta the
transmission probability, $|\Sct_\ell(\omega,M_i)|^2$ goes to zero.
This can
most easily be seen by recasting the scattering problem in terms of
the quantum
mechanics of a single particle moving in the presence of an effective
`potential':
\label\veff
\eq  \eqalign{
V_{\rm eff} &\equiv M^2_i \, \left( 1 - {\rs \over r} \right)
\left( 1 + { \ell(\ell +1) \over M_i^2 r^2 } + {\rs \over M_i^2
r^3} \right),
\cr
&\approx M^2_i \, \left( 1 - {\rs \over r} \right)
\left( 1 + { \ell^2 \over M_i^2 r^2 } \right). \cr} \eeq
This last, approximate, form has been simplified using
$\ell \gg 1$ and $M_i \rs \gg 1$. Classical evolution in
this potential simply predicts $|\Sct_\ell|^2 =1$ when
$\omega$ lies above the potential for all $r$, and $|\Sct_\ell|^2 = 0$
when $\omega$ is below the barrier for some $r$. For the above potential,
however, there is no barrier at all to escape for $\ell \le L\equiv
\sqrt{3} M_i
\rs$, since only for these $\ell$'s can the  centrifugal
contribution dominate
the gravitational attraction. For $\ell > L$, on the other hand,
$V_{\rm eff}$
has a maximum for $r = r_{\rm max} \gsim \rs$ that can reflect a
potentially
outgoing particle, and so transmission is forbidden for $\omega <
V_{\rm max}$.
But since the height of the barrier, $V_{\rm max} \sim \ell /M_i
\rs$, grows for
large $\ell$, reflection eventually becomes inevitable for
sufficiently large
$\ell$. Physically, particles with large $\ell$, but fixed
$\omega$, are not
sufficiently radially directed to escape to infinity once they try
to climb out
of the black hole's gravitational well. As a result the sum over
$\ell$ that
appears in $\Scf_i$ is eventually cut off for sufficiently large $\ell$.

We conclude, then, that at least for this regularization, the
contribution
of very-short-distance physics, at distances $\sim 1/\Lambda$,  to the
Hawking flux is exponentially suppressed by the large
ratio $\Lambda/\th$.

\section{Hawking Radiation and the Absence of Singularities}

The result of the last Section raises another question. We have
computed the Hawking radiation in a regulated theory having a
completely smooth two-point function. But it is also straightforward to show,
by
trivially extending the arguments of Ref. \fh\ to massive fields, that a
coincidence limit of the form $G(x,x') \sim 1/[4\pi^2 \sigma(x,x')] +
\hbox{(less singular)}$ --- where $\sigma(x,x')$ denotes the proper
separation between the points $x$ and $x'$ --- is required near $r=\rs$ in
order to produce the Hawking radiation. That is, in the approach of Ref.~\fh\
the Hawking flux is completely determined by the coefficient of this $1/\sigma$
singularity of the two-point function, $G(x,x')$, when the coincidence limit is
taken near the black hole event horizon. The question therefore is: How can a
nonzero flux be obtained using a regularized propagator which is smooth in the
coincidence limit? The present Section is devoted to the resolution of this
apparent contradiction.

The starting point for Ref.~\fh's analysis is the observation that
eq.~\coincidence\ allows us to compute the Hawking flux at a point
$(T,\Scr,\Theta,\Phi)$, at large distances from the black hole and
at late times, given knowledge of $G(x,x')$ in the neighbourhood of
this point. In ref.~\fh\ the two-point function at large distances
from the black hole, $G({X_1},{X_2})$, is related to its values on
an earlier spacelike hypersurface using the surface independence of
the Klein-Gordon inner product:
\label\innerproduct
\eq
(f,g) = \int_\Sigma f^* \lrderiv{\mu} \, g \; d\Sigma^\mu
\eeq
provided that the functions $f$ and $g$ satisfy the Klein-Gordon
equation. This leads to the following expression:
\label\fhintegral
\eq
G(X_1,X_2) =
\int_{\Sigma_\tau} \int_{\Sigma_\tau} d\Sigma_1^\mu d\Sigma_2^\nu \,
G(x_1,x_2)\, \lrderiv{1\mu} \, \lrderiv{2\nu} f(x_1)\, f^* (x_2) \, ,
\eeq
where both integrals are taken over the same timelike surface,
$\Sigma_\tau$,
which we may take to be a surface of constant $\tau = t + r^\star -
r$. The
measure for such a surface is $d\Sigma^\mu = n^\mu \,r^2
\sin^2\theta\,dr\,
d\theta \,d\phi$, with $n^\mu$ the unit normal to $\Sigma_\tau$.
Explicitly,
\eq
n\cdot \partial = \left( 1 + {r_s \over r} \right) {\partial \over
\partial
\tau} - {r_s \over r} {\partial \over \partial r} \, .
\eeq
The function $f(x)$, which appears in Eq.~\fhintegral\ is the particular
solution to the Klein-Gordon equation which satisfies the following
`initial'
conditions, which we choose to specify on a late-time constant-$t$
surface
which contains the point $X = (T,\Scr,\Theta,\Phi)$ at which the Hawking
flux is to be measured
\label\initialconditions
\eq \eqalign{
\Bigl. f(x) \Bigr|_{t = T} &= 0 , \cr
\Bigl. \partial_t f(x) \Bigr|_{t=T} &= \delta^3 (\vec{x} - \vec{X}). \cr}
\eeq
The vector symbol here denotes the three coordinates which specify
a point
on the surface $t=T$.

The Fredenhagen and Haag derivation \fh\ crucially relies on this
surface-independence of the Klein-Gordon inner product, $(f,g) = \int_\Sigma
f^* \lrderiv{\mu} \, g \; d\Sigma^\mu$, when the functions $f$ and $g$
satisfy
the Klein-Gordon equation. In eq.~\fhintegral, this is applied in
particular to the
two-point function, $G(x,x')$. The resolution of the apparent
paradox therefore
relies on the fact that a regulated propagator like $G_{\rm
reg}(x,x')$ does
{\it not} satisfy the Klein-Gordon equation, but rather satisfies a more
complicated higher-derivative equation of motion. The conserved
inner product
for this equation of motion also involves higher-derivative
corrections, and
these corrections are what generate the Hawking flux from a nonsingular
two-point function.

We next illustrate this argument with an explicit calculation.
Rather than
dealing with the cumbersome details of the five regulator fields
that are used
in the text, for clarity of presentation we instead present an
example which
uses just one regulator field. Consider, therefore, the following
two-point
function:
\eq \hat{G}(x,x') \equiv G_{m^2}(x,x') - G_{M^2}(x,x'),
\eeq
where $G_{m^2}(x,x')$ and $G_{M^2}(x,x')$ respectively
denote the two-point functions for a free scalar fields
of mass $m$ and $M \gg m$. Comparing to the short-distance
expansion of Ref.~\gilkey, shows that the coincidence limit
of $\hat{G}(x,x')$ is at worst $\sim \log \sigma(x,x')$, for
Schwarzschild spacetime. Even though this is less singular than
$1/\sigma(x,x')$, our goal here is to show that $\hat{G}(x,x')$
nevertheless produces a nonzero Hawking flux.

In order to apply the methods of Ref.~\fh, we must first find what
equation of motion $\hat{G}(x,x')$ satisfies, and then construct
the corresponding conserved `inner product' for this equation.
As is simple to check, the equation of motion is:
\eq
{1\over M^2 - m^2} \; \bigl( \Square - m^2 \bigr) \,
\bigl( \Square - M^2 ) \, \hat{G}(x,x') = 0 .
\eeq
The conserved `inner product' for two solutions, $f$ and $g$,
of this equation then is:
\eq
\eqalign{ [f,g]  & = - {M_+^2 \over M_-^2} \,
\int_\Sigma d\Sigma^\mu \; f^*
\lrderiv{\mu} g \, \cr
& \qquad + {1 \over M_-^2} \int_\Sigma \,d\Sigma^\mu \,
f^* \lrderiv{\mu} \Square \, g \; +
{1\over M_-^2} \int_\Sigma d\Sigma^\mu \, (\Square f^*)
\lrderiv{\mu} \, g, \cr} \eeq
where $M_\pm^2 \equiv M^2 \pm m^2$, and $\Sigma$ is a
spacelike surface. Clearly this expression approaches the
usual Klein-Gordon one in the
limit $M\to\infty$.

Using this expression to write the analogue of eq.~\fhintegral,
gives the following result
\eq
\hat{G}(X_1,X_2) = \sum_{j=1}^4 \hat{G}_j(X_1,X_2) ,
\eeq
where
\eq
\eqalign{
\hat{G}_1 (X_1,X_2) &= {1\over M_-^4} \,
\int_\Sigma \int_\Sigma d\Sigma_1^\mu
d\Sigma_2^\nu \, \hat{G}(x_1,x_2) \,
\lrderiv{1\mu} \, \lrderiv{2\nu} F(x_1) F^*(x_2) , \cr
\hat{G}_2(X_1,X_2) & = {1\over M_-^4} \,
\int_\Sigma \int_\Sigma d\Sigma_1^\mu
d\Sigma_2^\nu \, [\Square_1 \hat{G}(x_1,x_2)]
\lrderiv{1\mu} \, \lrderiv{2\nu}
f(x_1) F^*(x_2) ,\cr
\hat{G}_3(X_1,X_2) &= {1\over M_-^4} \,
\int_\Sigma \int_\Sigma d\Sigma_1^\mu
d\Sigma_2^\nu \, [\Square_2 \hat{G}(x_1,x_2)]
\lrderiv{1\mu} \, \lrderiv{2\nu}
F(x_1) f^*(x_2) ,\cr
\hat{G}_4(X_1,X_2) &= {1\over M_-^4} \,
\int_\Sigma \int_\Sigma d\Sigma_1^\mu
d\Sigma_2^\nu \,
[\Square_1 \Square_2 \hat{G}(x_1,x_2)] \lrderiv{1\mu} \,
\lrderiv{2\nu} f(x_1) f^*(x_2) .\cr}
\eeq
In these expressions $F(x) \equiv \Square f(x) - M_+^2 \, f(x)$.
The function $f(x)$ must also satisfy the following `initial conditions'
\eq
\eqalign{
\Bigl. f(x) \Bigr|_{t = T} &= 0 ,  \cr
\Bigl. \partial_t f(x) \Bigr|_{t = T} &= 0 ,  \cr
\Bigl. \Square f(x) \Bigr|_{t = T} &= 0 ,  \cr
\Bigl. \partial_t \Square f(x) \Bigr|_{t = T} &= M^2_- \,
\delta^3(x) ,  \cr}
\eeq
which play the role here of eq.~\initialconditions\ in the Klein-Gordon
case.

Fearsome as it looks, this initial-value problem can be solved, and
leads to
functions, $f^-(x)$, which are basically identical with those that
are found
for the Klein-Gordon case. In particular, their support becomes
infinitely
small as $(T-t) \to \infty$, requiring a coefficient function that
varies like
$1/\sigma(x_1,x_2)$ near the horizon. The new feature, though, is
that the
function that must be this singular involves not just
$\hat{G}(x_1,x_2)$, but
also its {\it derivatives}. It is these derivative terms that save
the day:
acting on $\hat{G}(x,x')$ they convert its $\log\sigma(x,x')$
behaviour into
the $1/\sigma(x,x')$ that is required for a nonzero result.

\section{Summary}

We have presented a derivation of the Hawking radiation within
a simple model for which the ultraviolet regularization has
been made explicit. This calculation permits the
regularization-dependence of the Hawking flux to be
explicitly displayed. It is found that the cutoff dependence
is exponentially small in the limit that $\Lambda/\th \gg 1$.
Since the Pauli-Villars regularization used is slice-independent,
this result agrees with what one would expect from the `nice-slice' argument
in favour of the irrelevance of the details of high-energy physics
on the prediction of Hawking radiation.

The computation scheme of Fredenhagen and Haag \fh\ is used
throughout, in which the Hawking radiation is directly related
to the coincident singularity of the two-point function as both
of its position arguments approach one another and the event
horizon. We show that there is no contradiction in this approach
between having a nonzero Hawking flux in the regulated theory,
even though the resulting regulated two-point function is
nonsingular in the coincidence limit.

\bigskip
\centerline{\bf Acknowledgments}
\bigskip

We thank Ted Jacobson for explaining to us the mysteries of
Ref.~\fh, and both Ted and Rob Myers for useful discussions.
This research was partially funded by funds from the
N.S.E.R.C.\ of Canada and les Fonds F.C.A.R.\ du Qu\'ebec.

\listrefs

\bye